\documentclass[twocolumn]{aastex631}
\usepackage{natbib}
\usepackage{booktabs}
\usepackage{multirow}

\newcommand{\nh} {$N_{\text{H}}$}
\newcommand{\chisq} {$\chi^{2}$}
\newcommand{\delchisq} {$\Delta \chi^{2}$}

\newcommand {\msun} {M$_{\odot}$}
\newcommand{\degree} {$^\circ$}

\shorttitle{SMC X-1 with NICER}
\shortauthors{Brumback et al.}

\begin{document}

\title{Constraining the evolution of the unstable accretion disk in SMC X-1 with NICER}

\correspondingauthor{McKinley C.\ Brumback}
\email{brumback@umich.edu}

\author[0000-0002-4024-6967]{McKinley C.\ Brumback}
\affiliation{Cahill Center for Astronomy and Astrophysics, California Institute of Technology, 1216 E California Blvd, Pasadena, CA 91125, USA}
\affiliation{Department of Astronomy, University of Michigan, 1085 S.\ University Ave. Ann Arbor, MI 48109 USA}

\author[0000-0003-3902-3915]{Georgios Vasilopoulos}
\affiliation{Department of Physics, National and Kapodistrian University of Athens, University Campus Zografos, GR 15783, Athens, Greece}

\author[0000-0001-7532-8359]{Joel B.\ Coley}
\affiliation{Department of Physics and Astronomy, Howard University, Washington, DC 20059, USA}
\affiliation{CRESST/Mail Code 661, Astroparticle Physics Laboratory, NASA Goddard Space Flight Center, Greenbelt, MD 20771, USA}

\author{Kristen Dage}
\affiliation{Department of Physics, McGill University, 3600 University Street, Montr\'eal, QC H3A 2T8, Canada}

\author{Jon M. Miller}
\affiliation{Department of Astronomy, University of Michigan, 1085 S.\ University Ave. Ann Arbor, MI 48109 USA}

\begin{abstract}
Neutron star high mass X-ray binaries with superorbital modulations in luminosity host warped inner accretion disks that occult the neutron star during precession. In SMC X-1, the instability in the warped disk geometry causes superorbital period ``excursions:" times of instability when the superorbital period decreases from its typical value of 55 days to $\sim$40 days. Disk instability makes SMC X-1 an ideal system in which to investigate the effects of variable disk geometry on the inner accretion flow. Using the high resolution spectral and timing capabilities of the Neutron Star Interior Composition Explorer (NICER) we examined the high state of four different superorbital cycles of SMC X-1 to search forchanges in spectral shape and connections to the unstable disk geometry. We performed pulse phase-averaged and phase-resolved spectroscopy to closely compare the changes in spectral shape and any cycle-to-cycle variations. While some parameters including the photon index and absorbing column density show slight variations with superorbital phase, these changes are most evident during the intermediate state of the supeorbital cycle. Few spectral changes are observed within the high state of the superorbital cycle, possibly indicating the disk instability does not significantly change SMC X-1's accretion process.
\end{abstract}

\section{Introduction}
Among high-mass X-ray binaries (HMXBs), super-giant X-ray binaries (SgXRBs) where the donor star overfills the Roche lobe offer the ideal laboratory for studying the interaction of the pulsar radiation with the accretion disk because the mass transfer via Roche-lobe overflow is relatively stable. The study of these systems has offered insight into the morphology of their accretion disk and its temporal stability. In particular, it is now well established that SgXRBs host warped disks (e.g.\ \citealt{ogilvie2001,hickoxvrtilek2005,brumback2020}), which cause superorbital modulation on timescales of 40-100 days as the warped disk edge partially obscures the compact object during its precession. Neutron star SgXRBs with superorbital modulation present a unique opportunity to probe the structure of the inner accretion disk because an observer simultaneously views the neutron star beam (dominated by hard X-ray radiation) and the soft, reprocessed emission from the inner accretion disk. Previous works by \cite{hickoxvrtilek2005}, \cite{brumback2020}, and \cite{Brumback2021} have demonstrated the success of using soft ($<1$ keV) pulse profiles to perform reverberation mapping in these sources, allowing them to model the geometry and kinematics of the warped accretion disks.

Perhaps the best systems for studying the ``soft-excess'' are SgXRBs in the Magellanic Clouds. These systems have known distances and suffer from little foreground absorption. In this respect SMC X-1 is potentially the best candidate for further study of warped accretion disks. SMC X-1 is a SgXB consisting of a 1.21(12) \msun\ NS with a $\sim 0.7$ s spin period orbiting an 18 \msun\ B0 supergiant companion every 3.9 days (\citealt{leong1971,schreier1972,webster1972,liller1973,falanga2015}).  
A radiation driven warp in the inner accretion disk, combined with the high orbital inclination ($i \sim 70$\degree ;\ \citealt{reynolds1993}), gives rise to a superorbital modulation in luminosity because the warped edge of the disk will partially obscure the pulsar during its precession (\citealt{pringle1996}). The disk precession causes three distinct superorbital states: the high state, which is characterised by maximum source flux, when the neutron star is unobscured by the disk, the low state, or time of minimum flux when the neutron star is occulted by the warped disk, and the intermediate state, which is the period of increasing or decreasing flux that marks the transition between high and low state. The period of this superorbital cycle varies between 40--60 days (e.g.\ \citealt{wojdowski1998,vrtilek2005}), which \cite{ogilvie2001} suggest arises from an instability in the accretion disk warp that causes the disk shape to vary between competing configurations. The superorbital cycle typically remains around 55 days in length, but during these periods of instability (known as ``excursions"), the cycle length can decrease down to 40 days (e.g. \citealt{hu2019,dage2019}).

The study of warped accretion disks in SgXRBs has potentially wider implications for a variety of systems. Super-orbital modulation of unknown origin is found in ultra-luminous X-ray systems (e.g.\ \citealt{vasilopoulos2021,gurpide2021,brightman2022}), which are a poorly understood class of X-ray binaries that host the brightest X-ray pulsars ($L_X>10^{39}$ erg/s or $>$10L Eddington; e.g.\ \citealt{king2023}). Warped accretion disks have also been observed in active galactic nuclei such as the Circinus galaxy and NGC 4945, making the geometry of these warped disk structures significant across mass and luminosity scales (e.g.\ \citealt{greenhill2003, martin2008}). 

Using the high resolution spectral and timing capabilities of the Neutron Star Interior Composition Explorer (NICER; \citealt{gendreau2016}), we examined the high state of different superorbital cycles of SMC X-1 to search for short term changes in spectral shape that might be caused by fluctuations in the unstable accretion disk geometry. Our data set, described in Section \ref{sec:obs}, consisted of 18 NICER observations covering four superorbital cycles. We performed pulse phase-averaged and pulse phase-resolved, and spectral analysis analysis to closely compare the spectral characteristics and any cycle-to-cycle variations (see Section \ref{sec:obs} for our analysis description and Section \ref{sec:results} for our results and discussion).

\begin{deluxetable*} {ccccccc}
\label{tab:obs} 
\tablecolumns{7}
\tablecaption{Description of NICER Observations} 
\tablewidth{0pt}
\tablehead{
\colhead{ObsID} & \colhead{Start Time (MJD)} & \colhead{Exposure (ks)} & \colhead{Orbital Phase} & \colhead{$F_{\text{0.6--10 keV}}$ (erg cm$^{-2}$ s$^{-1})$} & \colhead{Pulse Frequency (Hz)} }
\startdata
3638010101 & 59172.26 & 5.58 & 0.32 & (8.09$\pm$0.02)$\times10^{-10}$ & 1.432756(5) \\
3638010102 & 59173.10 & 5.15 & 0.54 & (8.84$\pm$0.02)$\times10^{-10}$ & 1.432652(5) \\
3638010103 & 59174.02 & 3.54 & 0.89 & (8.75$\pm$0.03)$\times10^{-10}$ & 1.43298(3) \\
3638010104 & 59175.57 & 0.12 & 0.17 & (8.2$\pm$0.1)$\times10^{-10}$ & 1.4324(3) \\
3638010105 & 59176.27 & 3.42 & 0.29 & (8.81$\pm$0.02)$\times10^{-10}$ & 1.432756(1) \\
3638010201 & 59216.25 & 8.17 & 0.62 & (4.08$\pm$0.01)$\times10^{-10}$ & 1.432809(1) \\
3638010202 & 59217.73 & 3.84 & 0.0 & N/A\tablenotemark{a} & N/A\tablenotemark{a} \\
3638010203 & 59218.52 & 5.70 & 0.21 & (6.64$\pm$0.02)$\times10^{-10}$ & 1.433064(3) \\
3638010204 & 59219.03 & 5.43 & 0.33 & (7.40$\pm$0.02)$\times10^{-10}$ & 1.432821(5) \\
3638010301 & 59368.64 & 0.55 & 0.78 & (8.87$\pm$0.07)$\times10^{-10}$ & 1.43355(4) \\
3638010302 & 59369.02 & 0.48 & 0.88 & N/A\tablenotemark{a} & N/A\tablenotemark{a} \\
3638010303 & 59370.25 & 6.00 & 0.19 & (9.15$\pm$0.03)$\times10^{-10}$ & 1.433382(7) \\
3638010304 & 59371.02 & 8.43 & 0.39 & (9.54$\pm$0.01)$\times10^{-10}$ & 1.432996(5) \\
3638010305 & 59372.00 & 2.07 & 0.64 & (9.23$\pm$0.03)$\times10^{-10}$ & 1.433183(2) \\
3638010306 & 59373.80 & 3.18 & 0.11 & (8.80$\pm$0.04)$\times10^{-10}$ & 1.433564(9) \\
3638010307 & 59373.99 & 0.77 & 0.15 & (9.32$\pm$0.07)$\times10^{-10}$ & 1.43344(7) \\
3638010401 & 59409.48 & 8.28 & 0.27 & (1.007$\pm$0.003)$\times10^{-9}$ & 1.433350(4) \\
3638010402 & 59409.99 & 10.99 & 0.40 & (1.026$\pm$0.002)$\times10^{-9}$ & 1.433151(1)
\enddata
\tablenotetext{a}{The orbital phases, fluxes, and light curves of this observation indicate that it occurred during either binary eclipse or a pre-eclipse dip. For this reason, this observation was not included in our analysis.}
\end{deluxetable*}

\section{Observations and Data Analysis} \label{sec:obs}

\subsection{NICER Observations} \label{sec:nicerobs}

Our data set consisted of 18 NICER observations taken between November 2020 and July 2021. The observations are grouped into four epochs (hereafter referred to as Epochs 1--4), with each epoch corresponding to a different superorbital cycle in either the bright or intermediate. Each epoch contains a varying number of NICER visits; Epoch 1 contains 5 visits, Epoch 2 contains 4 visits, Epoch 3 contains 7 visits, and Epoch 4 contains 2 visits. The observation information including obsID, exposure, and start time is contained in Table \ref{tab:obs}. For the sake of brevity, we refer to each observation by the last three digits of its obsID. In this format, the first digit corresponds to the epoch number and the second two digits refer to the visit number (e.g.\ Observation 302 is the second visit in the third epoch). 

\begin{figure*}
    \centering
    \includegraphics[scale=0.7]{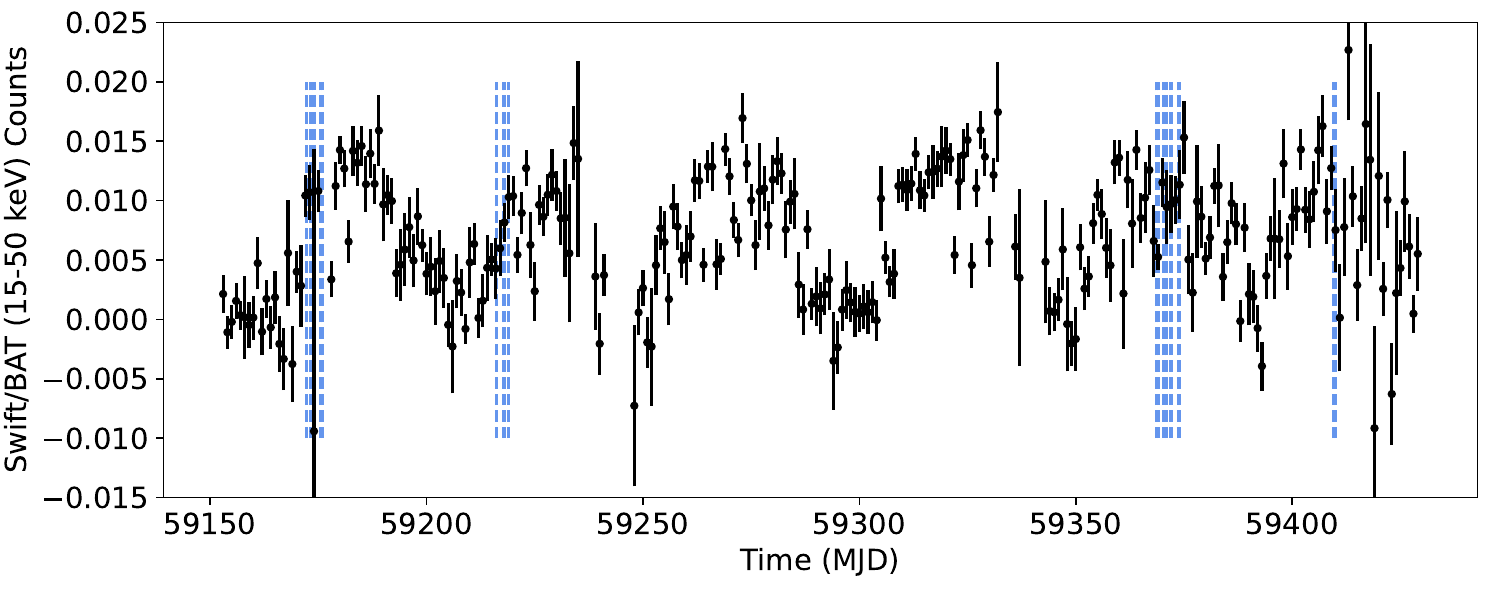}
    \caption{The Swift/BAT 15--50 keV light curve for SMC X-1 (black) showing the superorbital modulation around the time of our NICER observations (blue dashed lines). The observations occurred during the bright high state of four supeorbital cycles, referred to in this work as Epoch 1-4. Epoch 1 contains five NICER visits, Epoch 2 contains four NICER visits, Epoch 2 contains seven NICER visits, and Epoch 4 contains two NICER visits for a total of 18 observations.}
    \label{fig:longtermlc}
\end{figure*}

The observations in Epochs 1 and 2 occurred during the intermediate state of SMC X-1's superorbital cycle, as the source was increasing in brightness. Epochs 3 and 4 occured during the high state, when the source was at its brightest (see Figure \ref{fig:longtermlc}). We determined the length of the superorbital cycle containing each of our four epochs using the zero-crossing method (e.g.\ \citealt{smale2012}) with the Swift/BAT light curve binned to SMC X-1's orbital period. Taking the average of the cycle length found by the up-crossing and down-crossing times, we determined that the supeorbital cycles hosting Epochs 1--4 were approximately 46$\pm$3 days, 30$\pm$3 days, 44$\pm$3 days, and 46$\pm$3 days, respectively. From these cycle lengths and the excursion measurements of \cite{hu2023} it is clear that SMC X-1 was within a period of superorbital excursion during our observations (see also \citealt{hu2019,dage2019}).

The observations were reduced using the {\fontfamily{qcr}\selectfont nicerl2} pipeline within HEASoft v6.29c. To reduce contamination from background effects, we selected an underonly count range of 0--250 and an overonly count range of 0--1.5. Spectra and light curves were extracted using Xselect for each observation. Because SMC X-1 has a variety of known time dependent behaviors including binary eclipse and occasional pre-eclipse dips, we examined each light curve to determine the quality of the observation. Sixteen of the eighteen observations showed NICER count rates of approximately 250 counts s$^{-1}$ with little variability, consistent with viewing the source out of eclipse (e.g., \citealt{brumback2020}).

Because SMC X-1 is an eclipsing binary with known pre-eclipse dips, we examined the light curve of each observation to exclude times of low counts. Observations 202 and 302 showed low NICER count rates (less than 100 counts s$^{-1}$) and high variability consistent with eclipse and pre-eclipse dips. Calculating the orbital phase of these observations (using the \cite{falanga2015} ephemeris) indicated that Observation 202 occured at orbital phase 0 and Observation 302 occurred at orbital phase 0.88 (see Table \ref{tab:obs}). \cite{falanga2015} defines the eclipse in SMC X-1 to be between orbital phases 0.939(9) and 0.066(5), where the values in parenthesis represent the error on the last digit and the time of mid-eclipse is orbital phase 0. By these calculations, Observation 202 occured during binary eclipse. As seen in the light curves in \cite{brumback2022}, SMC X-1 sometimes exhibits pre-eclipse dips around orbital phase $\sim$0.9, therefore it is likely that the low count rates in Observation 302 are caused by a pre-eclipse dip. Because these observations fall during eclipse or a pre-eclipse dip, we excluded these two observations from the remainder of our analysis.

Further examination of the NICER light curves revealed that several observations (Observations 103, 201, 203, 303) had light curves that were predominantly bright but showed short intervals at the beginning or end of the observation in which the source decreased in flux. Since these intervals likely arise from the observation partially occurring during an eclipse or pre-eclipse dip, we filtered these observations by time in Xselect to remove the intervals of low count rates from our analysis.

For our final data set of sixteen observations, response files were generated with the {\fontfamily{qcr}\selectfont nicerarf} and {\fontfamily{qcr}\selectfont nicerrmf} tools. We generated background spectra using the {\fontfamily{qcr}\selectfont nibackgen3C50} model (\citealt{remillard2022}) and binned spectra using the FTOOL {\fontfamily{qcr}\selectfont ftgrouppha} with the optimum binning regime described in \cite{kaastra2016}.

\begin{figure*}
    \centering
    \includegraphics[scale=0.65]{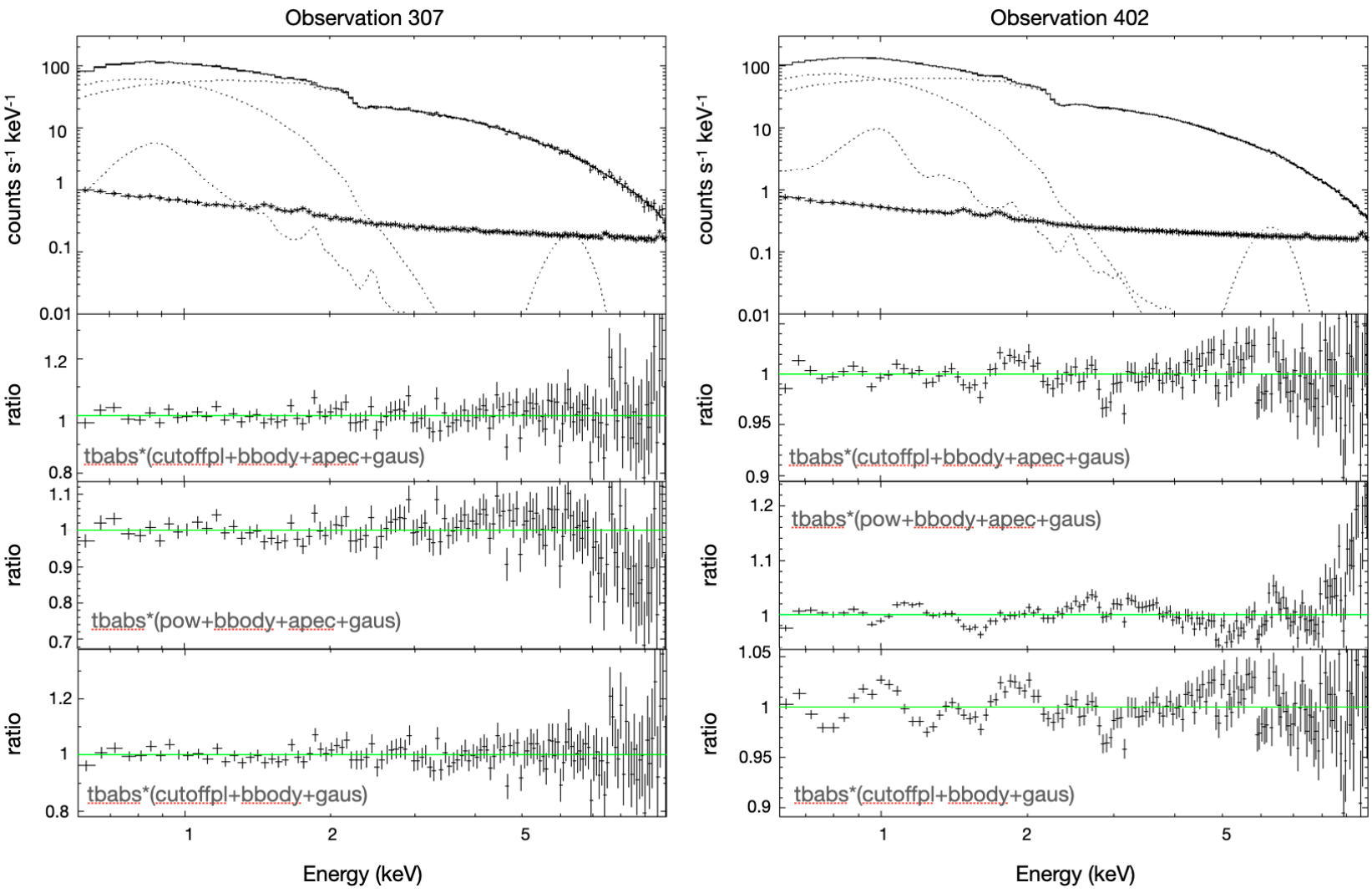}
    \caption{Phase-averaged spectra for Observations 307 (left) and Observations 402 (right) with the best fit spectral model {\fontfamily{qcr}\selectfont tbabs*(cutoffpl+blackbody+apec+gaus)} (top residual panel) The bottom two residual panels show alternative model components. The middle residual panels show the use of {\fontfamily{qcr}\selectfont pow} instead of {\fontfamily{qcr}\selectfont cutoffpl} which fails to accurately describe the power law tail in all spectra. The bottom residual panels show the best fit model without {\fontfamily{qcr}\selectfont apec}, the strength of which varies strongly between observations. This effect is illustrated here by the significant low energy residuals in Observation 402 (demonstrating a strong requirement for {\fontfamily{qcr}\selectfont apec}) compared to the minimal change in residuals in Observation 307. However, even in the case of Observation 307 the data prefer the presence of apec with a \delchisq of 8.6. }
    \label{fig:spectra}
\end{figure*}

\subsection{Phase-averaged spectroscopy}

We used Xspec to find the best fit phase-averaged spectral model (\citealt{arnaud1996}). We found that the soft spectral continuum was well described by a thermal blackbody, but prominent residuals below 2 keV remained in most spectra. Broad emission line features below 2 keV were also seen in previous XMM-Newton EPIC-pn observations of SMC X-1 (\citealt{brumback2020}). Chandra ACIS observations of SMC X-1 reveal these lines to be blends of Ne IX, Ne X, and O VIII (\citealt{vrtilek2001,vrtilek2005}). With our NICER spectra, we found that adding the collisionally-ionized diffuse gas model {\fontfamily{qcr}\selectfont apec} to our model provided a good fit to the soft continuum (see Fig. \ref{fig:spectra}). The {\fontfamily{qcr}\selectfont apec} model component varied in strength between different observations (as illustrated by Fig.\ \ref{fig:spectra}), and generally was more strongly required in the brighter spectra. Two of the faintest spectra, Observations 104 and 306, had poorly constrained {\fontfamily{qcr}\selectfont apec} parameters that did not significantly improve the \delchisq of the fit, and therefore we removed {\fontfamily{qcr}\selectfont apec} from the model in these two cases.

Additionally, we applied the photoelectric absorption model {\fontfamily{qcr}\selectfont tbabs} to model X-ray absorption between NICER and SMC X-1. We used abundances from \cite{wilms2000} and cross-sections from \cite{verner1996}. We attempted to add a second absorption component with abundances set to 20\% Solar values, as has been used in the past to represent absorption within the SMC itself, but found that this component was not required by the model (normalization set to zero).

The high energy continuum was described as a non-thermal continuum with a Gaussian emission line at 6.4 keV corresponding to Fe K$\alpha$. We trialed several different power law models to fit the non-thermal continuum above 2 keV including a power law ({\fontfamily{qcr}\selectfont pow}), a cutoff power law ({\fontfamily{qcr}\selectfont cutoffpl}), a power law with a high energy cutoff {\fontfamily{qcr}\selectfont pow*highecut}, and a negative and positive exponential (NPEX, \citealt{mihara1998}). The best fit was provided by {\fontfamily{qcr}\selectfont cutoffpl}, although the value of the high energy cutoff was poorly constrained due to NICER's upper energy limit of 12 keV and this affected the accuracy of our photon index values. Despite the poor constraints on the cutoff energy, the {\fontfamily{qcr}\selectfont cutoffpl} model was strongly preferred by the data over the pure power law model {\fontfamily{qcr}\selectfont pow}, which failed to correctly model the high energy emission (see Fig. \ref{fig:spectra}).

In order to better constrain the photon index, we attempted to fix the cutoff energy to the value given by archival NuSTAR data. To determine this, we used the two high state NuSTAR Observations 30202004002 and 30202004008 and fit them with a high energy X-ray continuum model of {\fontfamily{qcr}\selectfont constant*(tbabs*(cutoffpl+gaus))}. By taking the average of the cutoff energies from the two observations, we found a value of 11.06 keV for our NICER data set. However, fixing the cutoff energy to this value resulted in overestimation of the NICER data at high energies. To best fit the data while constraining the high energy cutoff, we therefore allowed the cutoff energy to vary around the NuSTAR value of 11.06 keV within the range of only 8--15 keV, and we recognize that our data are not sensitive to this parameter.

\subsubsection{Degeneracy between blackbody temperature and column density}

With a best fit continuum model of {\fontfamily{qcr}\selectfont tbabs*(cutoffpl+blackbody+apec+gaus)}, we discovered degeneracy between the absorbing column density and the blackbody temperature when both parameters were left free. We first used the HEASARC NH Calculator to estimate the column density towards SMC X-1 from the HI4PI survey, which suggested a density of $4.57\times10^{21}$ cm$^{-2}$ (\citealt{hi4pi2016}). This value was approximately four times larger than the average column density required by our model with both the \nh\ and blackbody temperature free ($\sim1\times10^{21}$ cm$^{-2}$). Fixing the \nh\ to the value from the HI4PI survey resulted in significantly larger \chisq\ and poorer fits.

Previous and ongoing NICER studies of SMC X-1 suggest that the blackbody temperature in SMC X-1 is a more constant spectral parameter than \nh\ (\citealt{dage2022}). In keeping with these findings, we limited the degeneracy between \nh\ and blackbody temperature by only allowing both parameters to be free in the brightest observation from each Epoch (Observations 105, 204, 304, and 402). In all other observations, we fixed the blackbody temperature to the value found in the brightest observation from that Epoch. This method still allowed  us to check for cycle-to-cycle variations in blackbody temperature while limiting the degeneracy with \nh. 

\begin{figure}[hb]
    \centering
    \includegraphics[scale=0.6]{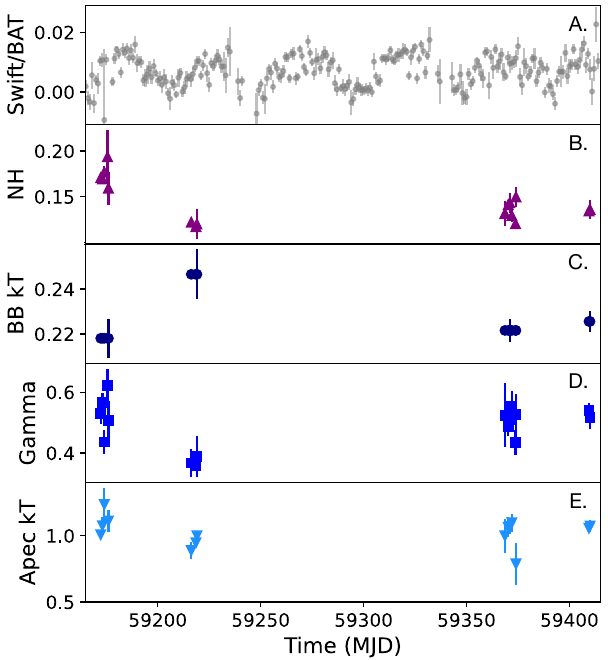}
    \caption{The key phase-averaged spectral parameters for our data set as a function of time. Panel A (grey points): The Swift/BAT 15--50 keV light curve for SMC X-1 (grey). Panel B (purple triangles): the absorbing column density (\nh). Panel C (navy circles): the blackbody temperature. Only one point for each epoch, corresponding to the brightest observation, shows the error bars. In the other observations, the temperature is fixed to the value from the brightest observation. Panel D (blue squares): the photon index. Panel E (light blue triangles): the plasma temperature from {\fontfamily{qcr}\selectfont apec}. Epochs 1 and 2, which occur during the intermediate state of the superorbital cycle, show the most significant differences in spectral parameters between cycles. Epochs 3 and 4 have values more consistent with each other. }
    \label{fig:phaseavgparams}
\end{figure}

After these tests, best fit phase averaged spectral model was {\fontfamily{qcr}\selectfont tbabs*(cutoffpl+blackbody+apec+ gaus)}, where the blackbody temperature was frozen to the value determined by the brightest observation in each Epoch. We limited the cutoff energy between 8--15 keV. The variations in key spectral parameters are plotted in Figure \ref{fig:phaseavgparams} and given in Table \ref{tab:phaseavgspecparams}. Fluxes for each observation, calculated from the {\fontfamily{qcr}\selectfont flux} command in Xspec, are shown in Table \ref{tab:obs}.

\subsection{Pulse profiles}
In preparation for timing analysis and making pulse profiles, the data were corrected to the Solar System barycenter using the FTOOL {\fontfamily{qcr}\selectfont barycorr}. The data were further corrected for the orbital motion of SMC X-1 using the ephemeris from \cite{falanga2015}, which gives the mid-eclipse time($T_{0}$) as 52846.688810(24) MJD, the orbital period at this epoch ($P_{\text{orb}}$) as 3.891923160(66) days, and the orbital decay ($\dot{P}_{\text{orb}}/P_{\text{orb}})$ as -3.541(2)$\times 10^{-6} \text{yr}^{-1}$.

\startlongtable
\begin{longrotatetable}
\begin{deluxetable*}{ccccccccccccc}
\label{tab:phaseavgspecparams}
\tablecaption{Phase-averaged spectral parameters}
\tablehead{\colhead{ObsID} & \colhead{$N_{\text{H}}$} & \colhead{Photon} & \colhead{Cutoff Energy\tablenotemark{a}} & \colhead{cutoffpl} & \colhead{kT$_{\text{blackbody}}$} & \colhead{blackbody} & \colhead{kT$_{\text{apec}}$} & \colhead{apec} & \colhead{Gaussian} & \colhead{d.o.f.} & \colhead{$\chi^{2}$} & \colhead{Reduced} \\ 
\colhead{} & \colhead{($10^{22} \text{cm}^{-2}$)} & \colhead{Index} & \colhead{(keV)} & \colhead{norm.} & \colhead{(keV)} & \colhead{norm.} & \colhead{(keV)} & \colhead{norm.} & \colhead{norm.} & \colhead{} & \colhead{} & \colhead{$\chi^{2}$} }
\startdata
101 & 0.171(5) & 0.53(3) & 10(1) & 0.0427(5) & 0.21(fixed & 0.00081(4) & 1.01(3) & 0.0028(4) & 0.0012(2) & 136 & 221.82 & 1.63 \\
102 & 0.169(5) & 0.57(3) & 11(1) & 0.0469(5) & 0.21(fixed) & 0.00089(5) & 1.08(7) & 0.0021(5) & 0.0009(2) & 136 & 184.09 & 1.35 \\
103 & 0.177(6) & 0.44(4) & 8.2(8) & 0.0458(6) & 0.21(fixed) & 0.00101(4) & 1.2(1) & 0.003(1) & 0.0006(3) & 131 & 146.24 & 1.12 \\
104 & 0.19(3) & 0.62(6) & 15(-7) & 0.042(3) & 0.21(fixed) & 0.0010(2) & \nodata & \nodata & 0.002(1) & 106 & 80.93 & 0.76 \\
105 & 0.16(2) & 0.51(6) & 11(2) & 0.044(1) & 0.218(9) & 0.00097(8) & 1.11(8) & 0.0028(7) & 0.0004(3) & 128 & 148.88 & 1.16 \\
201 & 0.122(5) & 0.37(5) & 8.4(9) & 0.0189(3) & 0.25(fixed) & 0.00035(2) & 0.89(7) & 0.0006(2) & 0.0004(1) & 129 & 158.42 & 1.23 \\
203 & 0.117(5) & 0.36(1) & 8.0(+2) & 0.0312(1) & 0.25(fixed) & 0.00055(2) & 0.95(3) & 0.0022(3) & 0.0008(2) & 137 & 184.95 & 1.35 \\
204 & 0.12(2) & 0.39(7) & 9(1) & 0.034(1) & 0.25(1) & 0.00062(4) & 1.00(4) & 0.0028(3) & 0.0007(2) & 128 & 209.22 & 1.63 \\
301 & 0.13(1) & 0.5(1) & 15(-3) & 0.045(2) & 0.222(fixed) & 0.0011(1) & 1.0(1) & 0.003(1) & 0.0014(7) & 115 & 123.57 & 1.07 \\
303 & 0.141(4) & 0.49(3) & 10(1) & 0.0444(5) & 0.222(fixed) & 0.00122(4) & 1.06(8) & 0.0021(5) & 0.001(2) & 139 & 265.33 & 1.91 \\
304 & 0.14(1) & 0.51(4) & 11(1) & 0.046(1) & 0.222(5) & 0.00122(5) & 1.07(6) & 0.0031(4) & 0.0010(2) & 141 & 271.92 & 1.93 \\
305 & 0.129(7) & 0.55(5) & 15(-2) & 0.0459(9) & 0.222(fixed) & 0.00106(7) & 1.10(7) & 0.0037(9) & 0.0008(4) & 130 & 154.64 & 1.19 \\
306 & 0.120(5) & 0.43(4) & 10(1) & 0.0413(6) & 0.222(fixed) & 0.00096(4) & \nodata & \nodata & 0.0005(3) & 133 & 223.10 & 1.68 \\
307 & 0.15(1) & 0.53(7) & 15(-3) & 0.046(1) & 0.222(fixed) & 0.0011(1) & 0.8(2) & 0.0016(9) & 0.0007(6) & 120 & 118.78 & 0.99 \\
401 & 0.134(3) & 0.54(3) & 13(1) & 0.0488(5) & 0.226(fixed) & 0.00118(4) & 1.06(3) & 0.0038(4) & 0.0009(2) & 141 & 228.20 & 1.62 \\
402 & 0.14(1) & 0.52(4) & 12(1) & 0.050(1) & 0.226(4) & 0.00124(5) & 1.07(4) & 0.0036(3) & 0.0009(2) & 142 & 316.03 & 2.23 \\
\enddata
\tablecomments{Phase averaged spectral parameters for the base model of {\fontfamily{qcr}\selectfont tbabs*(cutoffpl+blackbody+apec+gaus)}. In Observations 104 and 306, the {\fontfamily{qcr}\selectfont apec)} component was poorly constrained, with normalizations of effectively 0, so in these two observations we removed the {\fontfamily{qcr}\selectfont apec)} model component. Astericks (*) denote the brightest observation in each Epoch.}
\tablenotetext{a}{The cutoff energy is limited between 8--15 keV to provide constraints for this parameter. When the cutoff energy fits to the lower limit (8 keV) then it has only a positive error value, whereas when it fits to the upper limit (15 keV) it has only a negative error value. We emphasize that this parameter is poorly constrained in the NICER energy bandpass.}
\end{deluxetable*}
\end{longrotatetable}

For each observation, we used the HENDRICS (\citealt{hendrics}) and Stingray (\citealt{stingray}) software packages to perform the timing analysis. First, we confirmed the presence of pulsations in each of our observations by creating power density spectra in HENDRICS, which showed the presence of strong pulsations at approximately 1.4 Hz. We used the HENDRICS folding search to determine the frequency of the pulsations and the frequency first derivative. In all observations, the spin frequency first derivative was consistent with zero. We calculated the uncertainty of the spin frequency as the change in observed pulse phase over the course of each observation ($\delta\nu = \delta\text{phase} / \delta\text{time}$). To calculate the change in pulse phase we used Xselect to extract pulse profiles from short time intervals at the beginning and end of each observation and measured the difference in phase between the profiles. We present the pulse frequency in Table \ref{tab:obs} where the number in parentheses is the uncertainty on the last digit.

\begin{figure*}
    \centering
    \includegraphics[scale=1.0]{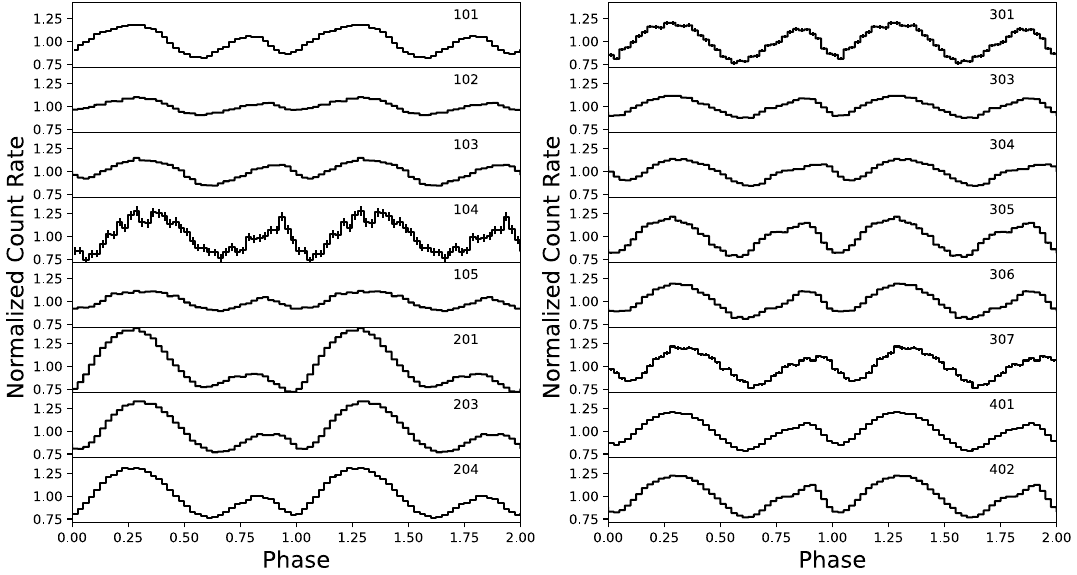}
    \caption{Pulse profiles for our data set in the full NICER bandpass. Pulse profiles have been shifted to align the primary pulse peak and all plotted with the same y-axis scale. All profiles show the double peaked structure characteristic of SMC X-1, but the relative strength of the pulse peaks varies between observations.}
    \label{fig:allpp}
\end{figure*}

The entire light curve was used to make the final pulse profiles for each observation, except in the case of light curves that had been previously filtered to remove low count rate intervals as described in section \ref{sec:nicerobs}. Final pulse profiles in the entire NICER bandpass were made using the Stingray epoch folding tool, the pulse frequency described above, and the start time of the observation as the starting epoch. The pulse profiles are shown in Figure (\ref{fig:allpp}).

\subsection{Pulse phase-resolved spectroscopy}
We employed pulse phase-resolved spectroscopy in order to examine how spectral parameters vary with spin phase. Each pulse profile was separated into four phase bins corresponding to the primary and secondary peaks of the pulse profile and the two troughs between them. We filtered by these spin phase bins in Xselect to create four pulse phase-resolved spectra.

As expected, the pulse phase-resolved spectra are of weaker signal to noise than the phase-averaged spectra. Even for the spectra corresponding to the primary and secondary pulse peaks, where the signal is highest, we did not detect emission lines present in the phase-averaged spectra. Therefore, we removed the {\fontfamily{qcr}\selectfont apec} and {\fontfamily{qcr}\selectfont gaussian} model components corresponding and fit all pulse phase-resolved spectra with the simplified {\fontfamily{qcr}\selectfont tbabs*(cutoffpl+bbody)} model. To assist our spectral modeling, we also froze the blackbody temperature and the cutoff energy to their phase-averaged values for each pulse phase-resolved spectrum. The resulting spectral model allowed us to test how the \nh, photon index, power law normalization, and blackbody normalization varied with pulse phase.

\begin{figure*}
    \centering
    \includegraphics[scale=0.8]{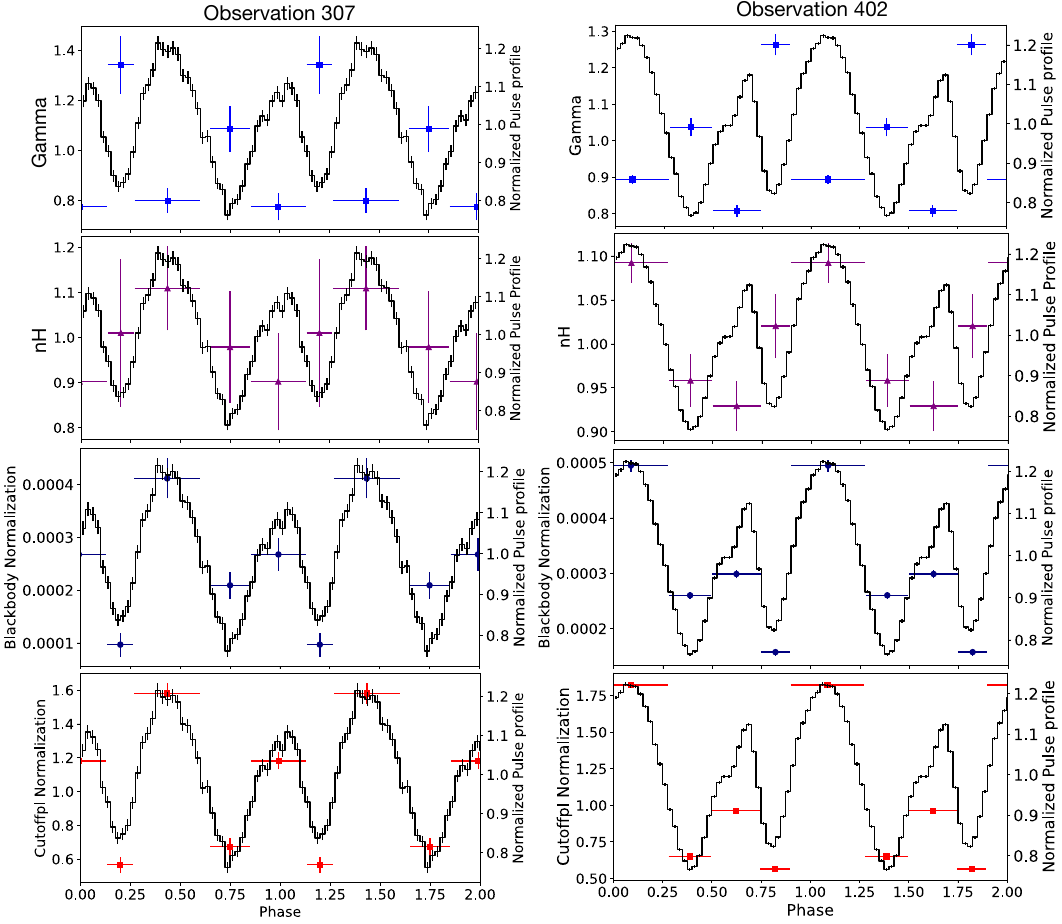}
    \caption{Pulse phase-resolved spectral parameters for Observations 307 and 402 as an example of the behavior seen in the full data set. In all panels, the solid black line is the full bandpass NICER pulse profile. Both the pulse profile and the parameter values have been normalized against their mean values to enable plotting on a uniform scale. First panel (blue squares): The photon index shows strong variation with pulse phase, increasing in strength during the pulse trough and decreasing during the pulse peak. Second panel (purple triangles): The \nh\ does not show changes with pulse phase in some observations, like 307, while in others it does show changes as in Observation 402. Third panel (navy circles) the normalization of the blackbody component follows the shape of the pulse profile. Fourth panel (red squares): The normalization of the power law component follows the shape of the pulse profile. }
    \label{fig:phaseresparams}
\end{figure*}

For the sake of brevity, we plot pulse phase-resolved spectral parameters from Observations 307 and 402, which illustrate the types of variations seen in our data, in Figure \ref{fig:phaseresparams}.

\section{Results and Discussion} \label{sec:results}
\subsection{Spectral variations with superorbital cycle}
Our spectral analysis suggests that SMC X-1 does exhibit some changes in spectral shape between superorbital cycles, but that these changes may depend more on supeorbital phase than cycle length. The greatest difference can be seen between the \nh, blackbody temperature, and photon indices of Epochs 1 and 2, which both take place during SMC X-1's intermediate state as it is increasing in brightness (Figure \ref{fig:phaseavgparams}). Epochs 3 and 4, which occur during the brightest parts of their superorbital cycles when the source flux is not changing as rapidly, show spectral parameters that are generally consistent with each other.

The observations in Epoch 1 and Epoch 3 show the greatest overall distribution in phase-averaged spectral parameters (Figure \ref{fig:phaseavgparams}). This is potentially caused by the varying exposure times of the observations in these epochs; both epochs contain long exposures (greater than 2 ks) and short exposures (less than 1 ks), resulting in a wide range in spectral quality.

Our 16 phase-averaged spectral fits produced an average photon index of 0.49 and an average \nh\ of 1.4 $\times 10^{21} \text{cm}^{-2}$. These values are both smaller than the average photon index and \nh\ found by \cite{dage2022} (0.69 and 2.48 $\times 10^{21} \text{cm}^{-2}$ , respectively), whose NICER observations of SMC X-1 occurred later during the 2020-2022 superorbital excursion. The difference in these values could arise from differences in the spectral modeling; \cite{dage2022} restricted the cutoff energy in their fits to vary between 12 and 20 keV and we used a range of 8 to 15 keV. The cutoff energy is poorly constrained in both our data set (see large fluctuations in Table \ref{tab:phaseavgspecparams}) and in \cite{dage2022} (see their Section 3.1). Additionally, \cite{dage2022} used Gaussian emission lines to model the soft emission line blends where we used the collisionally ionized plasma model {\fontfamily{qcr}\selectfont apec}.

To test whether the difference between our average \nh and photon index values compared to \cite{dage2022} was a result of differing spectral model, we applied their same spectral model to our 16 spectra. When using the \cite{dage2022} model on our data set, the mean \nh was 1.85$\times 10^{21} \text{cm}^{-2}$ and the mean photon index was 0.59. These values are closer to the values of \cite{dage2022} but not in full agreement, which implies that our differing spectral model was only partly responsible for the discrepancy.

One possible explanation of this discrepancy is spectral selection criteria. Our data set comprises of both superorbital high and intermediate state spectra that vary in exposure time and overall data quality. \cite{dage2022} modeled only high quality spectra from the superorbital high state. Indeed, when we examine our \nh and photon index values from our brightest observations fit with the \cite{dage2022} spectral model, we find individual photon indices in better agreement with their values. An alternative explanation for the difference in averaged photon index and \nh\ values between our work and \cite{dage2022} is that SMC X-1 has undergone an intrinsic spectral change as it progresses through its excursion and period of disk instability. A direct analysis of all the NICER data of SMC X-1 during excursion would determine whether or not this is the case, but we defer such an analysis for future work.

Despite the difference in average value between our continuum parameters and those of \cite{dage2022}, our analysis does confirm their finding that the spectral shape of the bright state observations are general consistent. \cite{dage2022} modeled 11 spectra from the bright state of SMC X-1's superorbital cycle between April 2021 and January 2022 and found very little fluctuation in their spectral parameters over this time period (see Table 2 of \citealt{dage2022}). We see similar consistency within our Epoch 3 and Epoch 4 bright state data. The overall consistency of SMC X-1's bright state spectrum with time appears to imply that the inner accretion flow may be largely unaffected by changes to the accretion disk geometry during excursion.

\subsection{Pulse profiles and pulse phase-resolved spectroscopy}
In Figure \ref{fig:allpp} we show the NICER pulse profiles for the 16 observations that we analyzed. Although all the pulse profiles have double peaked shape characteristic of SMC X-1, both the strength of the pulsations and the relative strengths of the pulse peaks vary with time. Similar to our spectral results, changes in pulse profile shape are small within a single epoch and larger between epochs, although overall changes in pulse strength can be seen as exposure times vary, especially in Epoch 1.

Our spectral analysis indicated very little change between the spectral shape of Epochs 3 and 4, and our timing analysis provides further evidence of the similarity between high state properties. Our pulse profiles from Epochs 3 and 4 (Fig.\ \ref{fig:allpp}, right panel) show the least variation both in shape and strength. 

In our pulse phase-resolved spectral analysis, we filtered each observation by pulse phase into four bins corresponding to the two peaks and two troughs of the pulse profile. We used a simplified spectral model of {\fontfamily{qcr}\selectfont tbabs*(bbody+cutoffpl)} with the blackbody temperature and cutoff energy fixed to their phase-averaged values to reduce degeneracy in the pulse phase-resolved spectra.

In all observations, the photon index showed a clear anti-correlation with pulse profile (Figure \ref{fig:phaseresparams}, top panels); the photon index increased during the troughs of the pulse profile and decreased during the pulse peaks. We interpret this as viewing the accretion column during the pulse peaks, which results in a harder X-ray spectrum.

The pulse phase-resolved results for the \nh\ are more challenging to interpret. For some observations (as illustrated by Observation 307 in Figure \ref{fig:phaseresparams}), the \nh\ is consistent with being constant across pulse phase. In other observations (as illustrated by Observation 402), significant changes in \nh\ are seen with pulse phase, but these changes do not show a clear trend with pulse phase. The fact that we needed to freeze the blackbody temperature in order to obtain meaningful information from the \nh\ means that some of these variations could be due to changes in blackbody temperature, such as seeing the neutron star's hot spot rotate in and out of the line of sight. Alternatively, the changes in \nh\ could be more sensitive to binary phase and the position of the stellar wind. Ultimately, the variations in \nh\ with spin phase are not sufficiently clear to draw physical conclusions.

Finally, the blackbody and power law normalization values follow the shape of the overall pulse profile as it varies in flux, which is consistent with our expectations for brighter spectra from the pulse peaks.

\section{Conclusions}
In this work we analyzed 16 NICER observations of the SgXB SMC X-1. These observations spanned four different superorbital cycles, with Epochs 1 and 2 occurring during the intermediate state of the superorbital cycle and Epochs 3 and 4 occurring during the bright state. We performed both phase-averaged and pulse phase-resolved spectral analysis and also examined the pulse profiles for each of our observations. We summarize our primary findings here.

The pulse phase-averaged spectral parameters from the superorbital high state (Epochs 3 and 4) show little difference from one another. This indicates that the soft X-ray spectrum, which tracks the inner accretion flow, is likely insensitive to changes in superorbital cycle length and accretion disk geometry. This result is supported by our timing analysis, which indicates that the pulse profiles from Epochs 3 and 4 show little variation in shape and strength.

The pulse phase-averaged spectral parameters from the intermediate state of the superorbital cycle (Epochs 1 and 2) show more variation between cycles and significant differences in pulse profile shape. This indicates that the soft X-ray spectrum of the superorbital intermediate state is less consistent, possibly due to fact that, during this part of the cycle, we witness the neutron star emerging from behind the accretion disk so our view of the neutron star is rapidly changing. Our pulse phase-resolved spectroscopy also showed a strong anti-correlation between pulse peak and photon index, however variations in \nh\ are less clear.

\begin{acknowledgements}
We would like to thank the anonymous referee for comments and suggestions that strengthened the analysis and manuscript. MCB acknowledges support from NASA grant 80NSSC21K0263. GV acknowledges support by H.F.R.I. through the project ASTRAPE (Project ID 7802).
 KCD acknowledges fellowship funding from Fonds de Recherche du Qu\'ebec $-$ Nature et Technologies, Bourses de recherche postdoctorale B3X no. 319864.  Joel B. Coley acknowledges support by NASA award number 80GSFC17M0002.
\end{acknowledgements}

\software{HEAsoft (v6.29c; HEASARC 2014), NuSTARDAS, Stingray (\citealt{stingray}), HENDRICS (\citealt{hendrics}), MaLTPyNT (\citealt{maltpynt})}

\pagebreak

\bibliography{main.bib}{}
\bibliographystyle{aasjournal}

\end{document}